\documentclass[%
reprint,%
amssymb, amsmath,%
aip,cha,rsi%
]{revtex4-1}

\usepackage{docs}%
\usepackage{bm}%
\usepackage[colorlinks=true,linkcolor=black]{hyperref}%
\usepackage{amsmath}
\usepackage{graphicx}
\usepackage{cancel}
\usepackage{subfigure}
\usepackage{mathrsfs}
\usepackage{textcomp}
\usepackage{upgreek}
\usepackage{braket}
\expandafter\ifx\csname package@font\endcsname\relax\else
\expandafter\expandafter
\expandafter\usepackage
\expandafter\expandafter
\expandafter{\csname package@font\endcsname}%
\fi
\begin{document}
	
\title{Sensitivity-Enhanced Magnetometry Using Nitrogen-Vacancy Ensembles via Adaptively Complete Transitions Overlapping}
	
\author{Bao Chen}%
\affiliation{School of Physics, Hefei University of Technology, Hefei, Anhui 230009, China}

\author{Bing Chen}%
 \altaffiliation{The authors to whom correspondence should be addressed: bingchenphysics@hfut.edu.cn and nyxu@zhejianglab.edu.cn}
\affiliation{School of Physics, Hefei University of Technology, Hefei, Anhui 230009, China}%

\author{Xinyi Zhu}%
\affiliation{School of Physics, Hefei University of Technology, Hefei, Anhui 230009, China}%

\author{Jingwei Fan}%
\affiliation{School of Physics, Hefei University of Technology, Hefei, Anhui 230009, China}%

\author{Zhifei Yu}%
\affiliation{School of Physics, Hefei University of Technology, Hefei, Anhui 230009, China}%

\author{Peng Qian}%
\affiliation{School of Physics, Hefei University of Technology, Hefei, Anhui 230009, China}%

\author{Nanyang Xu}%
 \altaffiliation{The authors to whom correspondence should be addressed: bingchenphysics@hfut.edu.cn and nyxu@zhejianglab.edu.cn}
\affiliation{Research Center for Quantum Sensing, Zhejiang Lab, Hangzhou, 311000, China}%
	
\begin{abstract}	
Nitrogen-vacancy (NV) centers in diamond are suitable sensors of high-sensitivity magnetometry which have attracted much interest in recent years. Here, we demonstrate sensitivity-enhanced ensembles magnetometry via adaptively complete transitions overlapping with a bias magnetic field equally projecting onto all existing NV orientations. Under such conditions, the spin transitions corresponding to different NV orientations are completely overlapped which will bring about an obviously improved photoluminescence contrast. And we further introduce particle swarm optimization into the calibration process to generate this bias magnetic field automatically and adaptively using computer-controlled Helmholtz coils. By applying this technique, we realize an approximate 1.5 times enhancement and reach the magnetic field sensitivity of $\rm855\ pT/\sqrt{Hz}$ for a completely overlapped transitions compared to $\rm 1.33\ nT/\sqrt{\rm Hz}$ for a separate transition on continuous-wave magnetometry. Our approach can be conveniently applied to direction-fixed magnetic sensing and obtain the potentially maximum sensitivity of ensemble-NV magnetometry.     
\end{abstract}

\maketitle
	
\section{INTRODUCTION}
	Nitrogen-vacancy (NV) centers have been proved to be a promising quantum sensing platform for various physical quantities, such as magnetic field \cite{maze2008nanoscale,hall2012high,jakobi2017measuring,shim2022multiplexed}, electric field \cite{dolde2011electric,li2020nanoscale}, and temperature \cite{kucsko2013nanometre,neumann2013high}. 
	Especially, ensembles of NV centers were proposed for magnetometers with theoretically high sensitivity up to femtotesla scale as well as millimeter-level spatial resolution \cite{taylor2008high}. Compared to conventional magnetic sensing technologies such as superconducting quantum interference devices (SQUID) \cite{doi:10.1126/science.7361105} and atomic-vapor magnetometry \cite{kitching2018chip}, ensemble-NV-diamond magnetometry shows 
	a series of advantages including room temperature and pressure working requirement, biocompatibility, and simplicity of the experimental equipment. Hence, it has already been applied into biological active neuron magnetic sensing \cite{barry2016optical}, magnetic imaging on integrated circuits \cite{turner2020magnetic} or emerging materials \cite{ku2020imaging}, etc.   
	
\begin{figure*}
		\includegraphics{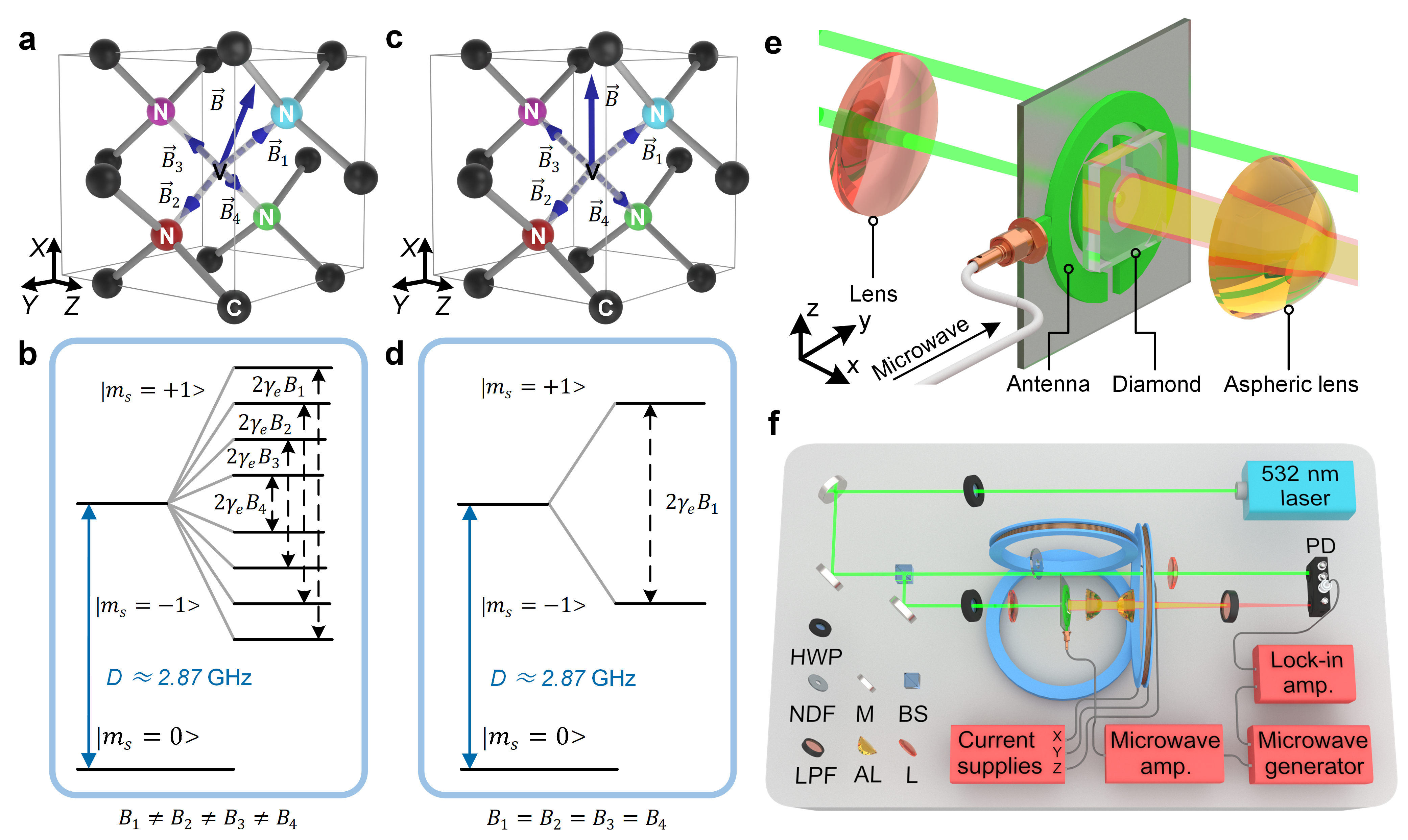}
		\caption{\label{fig:setup} (a) The structure of NV centers in diamond including the bias magnetic field $\vec{B}$ (solid line with arrow) and its vector projections (dashed lines with arrows), where $\vec{B}=\vec{B}_{1}-\vec{B}_{2}+\vec{B}_{3}-\vec{B}_{4}$. Four possible positions of a substitutional nitrogen for a fixed vacancy are marked with different colors. The coordinates in this figure and (c) have been rotated to be consistent with that in lab frame. (b) The ground-state electronic spin levels of NV centers and their quadruple Zeeman splittings of $\ket{m_{s}=\pm1}$ state corresponding to the bias magnetic field in (a). (c) The bias magnetic field whose vector projections are equal in their magnitudes. (d) The complete overlapped transitions corresponding to the bias magnetic field in (c). (e) Enlarged view of the core of ensemble-NV magnetometry. (f) Schematic diagram of the overall experimental setup. HWP, half-wave plate; NDF, neutral density filter; M, mirror; BS, beamsplitter; LPF, long pass filter; L, lens; AL, aspheric lenses; PD, photodetector. For simplification, three current supplies are integrated as one and only half of the three pairs of Helmholtz coils are shown here. Part of 3D models in (e) and (f) are from Ref. \cite{thorlabs} with permission.}
	\end{figure*}

	NV centers are color centers existing in diamond. Each NV center holds one of four orientations according to the C$_3\nu$ symmetry of the lattice as shown in Fig.\ref{fig:setup}(a). And its tetrahedral bond angle angular can be obtained as $\rm \theta_{tet}=109.47^\circ$ from geometric calculations \cite{chen2020calibration}. For a typical ensemble of NV centers , these different orientations are approximately equivalent in quantity. The ground state ($^3A_2$) of the NV center is a spin triplet with a zero-field splitting of 2.87 GHz between $\ket{m_{S}=0}$ and $\ket{m_{S}=\pm1}$ sub-levels, see Fig.\ref{fig:setup}(b). In an external magnetic field $\vec{B}$, $\ket{m_{S}=\pm1}$ sub-levels separate and shift proportional to $\gamma_{e}B_{i}$ due to the Zeeman effect, where $\rm \gamma_{e}\approx \rm{28}\ MHz/mT$ is the NV electron gyromagnetic ratio and $B_{i},\ i=\left\lbrace1,2,3,4 \right\rbrace$ are scalar projections of $\vec{B}$ onto the NV center axes. Thus, the external magnetic field can be determined by optically detected magnetic resonance (ODMR) spectrum measurement. This provides a basic method for magnetometry using an ensemble of NV centers. In contrast of the methods depended on pulsed laser, continuous-wave (cw) ODMR protocol is favored for its simple implementation and considered to potentially yielding a sensitivity similar to Ramsey protocols for near dc fields \cite{barry2020sensitivity}. 
	
	However, noticing the extant gap between the already achieved sensitivity at low frequency and its theoretical limitation, researchers have been dedicating diverse methods to enhance the magnetic-field sensitivity based on cw-ODMR. In the absence of sequences, these methods primarily exist in excitation enhancement \cite{ahmadi2017pump,clevenson2015broadband,PhysRevB.97.024105}, fluorescence collection efficiency improvement \cite{wolf2015subpicotesla,yu2020enhanced}, diamond sample optimization \cite{barry2016optical,edmonds2020generation}, magnetic flux concentrators \cite{fescenko2020diamond,xie2021hybrid}, etc. In most of current works, permanent magnets or Helmholtz coils provide the bias field, whose direction is approximately along $\left[111\right]$ crystallographic orientation, corresponding to only one NV axis used for magnetic sensing. In practice, we expect as many NV axes as to be involved in the sensing process, which can provide higher photoluminescence (PL) contrast to improve the conversion ratio from voltage to magnetic flux density.

	Here, we overcome the difficulty in applying the optimal bias field and realize a significant enhancement of sensitivity by overlapping all the NV ground-state spin transitions. Based on the particle swarm optimization (PSO) strategy, the direction of the optimal field can be automatically determined during the calibration procedure. Thus, the diamond sensor can be adaptively placed close to the sample, resulting in an increased ratio of output voltage to the magnetic field. Finally, with this strategy applied to our magnetometer, its magnetic sensitivity is increased from $\rm 1.33\ nT/\sqrt{\rm Hz}$ to $\rm855 \ pT/\sqrt{\rm Hz}$, yielding a 1.5 times enhancement.
	
\section{EXPERIMENT METHOD AND SETUP}	
	For cw-ODMR magnetometry based on NV ensembles, frequency modulation microwave is commonly utilized with lock-in method to extract the magnetic signals from the naturally ambient noise, i.e., the $1/f$ noise. The microwave frequency is tuned to the maximal slope point in the lock-in amplified ODMR spectrum as the magnetometer's working point, where the discrepancy of the fluorescence brought by minimum detectable magnetic field change is maximum. And its shot-noise-limited sensitivity is given as \cite{barry2016optical,dreau2011avoiding}: 
	\begin{equation}
		\label{eq:sens}
		\eta_{B}\approx\mathcal{P}_{\mathcal{F}}\frac{h}{g_{e}\mu_{B}}\frac{\rm \Delta\nu}{\mathcal{C}\sqrt{\mathcal{R}}},
	\end{equation}
	where $\mathcal{R}$ is the rate of detected photons, $\Delta \nu$ is the full width at half maximum (FWHM) resonance linewidth, $\mathcal{C}$ is the PL contrast, and $\mathcal{P}_{\mathcal{F}}$ is the coefficient calculated from the specific lineshape of the spin resonance which is $4/3\sqrt{3}$ for a Lorentzian lineshape. Experimentally, the enhancement of sensitivity is mapped onto the promotion of the ratio of contrast and linewidth $\mathcal{C}/\Delta \nu$ as well as the decrease of the overall noise level. Considering the hyperfine splitting of $^{14}$N nuclear spin and ignoring both the optical and microwave broadening, a typical Lorentzian profile including four main peaks can be shown as: 
	\small{
		\begin{equation}
			\label{eq:Lorentzian}
			F\left( \omega \right )=F_{0}(1-\sum_{i=1}^{4}\sum_{j=-1}^{1}\mathcal{C}\frac{(\Delta \nu/2)^{2} }{(\Delta\nu/2)^{2}+(\omega+j\Delta\omega_{\rm HF}-g_{e}\mu _{B} B_{i} )^2}),
	\end{equation}}
	where $F_{0}$ is the background fluorescence, $\Delta\omega_{\rm HF}=2.16$ MHz is the frequency shift of hyperfine splitting, $g_{e}\approx2.003$ is the negatively charged NV center's electron-spin $g$ factor and $\mu_{B}$ is the Bohr magneton. The other four peaks which are symmetrical to $\rm D\approx2.87\ GHz$ are not considered in here for simplification. And if the direction of bias magnetic field is coincide with the angular bisector in any two NV axes, we will have $B_{1}=B_{2}=B_{3}=B_{4}$. Under this condition, Eq.\,\ref{eq:Lorentzian} can be rewritten as:
	\begin{equation}
		\label{eq:Lorentzian_overlapped}
		F\left( \omega \right )=F_{0}(1-\sum_{j=-1}^{1}4\mathcal{C}\frac{(\Delta \nu/2)^{2} }{(\Delta\nu/2)^{2}+(\omega+j\Delta\omega_{\rm HF}-g_{e}\mu _{B}B_{1} )^2}).
	\end{equation}
	Assuming an ideal power uniformity of the frequency sweeping window across all the features, the PL contrast is improved by fourfold without worsening the linewidth and the sensitivity will get corresponding enhancement. This strategy equivalently increases the number of NV centers involved in the magnetic sensing. Meanwhile, no extra $\rm ^{13}C-NV$ interaction is introduced which will broaden the NV spin resonance via limiting the coherence time $T_{2}^{*}$. However, the magnetic field under test is also reduced by multiplying the factor $\rm cos(\theta_{tet}/2)$ due to the projection measurement inherently. As a result, reciprocal of the magnetic field sensitivity is expected to be enhanced by $\rm 4cos(\theta_{tet}/2)\approx2.3$ times.   
	
	To obtain the practical sensitivity enhancement, the experimental setup is constructed based on a single-crystal diamond (3$\times$3$\times$1.5 mm$^{3}$, Element Six). The diamond which is excited by focused 532-nm laser and microwave is glued on a printed circuit board patterned with a double split-ring resonator, as shown in Fig.\,\ref{fig:setup}(e). A balanced amplified photodetector is utilized for noise-reduced PL detection by common mode rejection. Similar uses are shown in Ref. \cite{patel2020subnanotesla,clevenson2015broadband,fescenko2020diamond,doi:10.1063/1.5095241}. The sample is placed in the central zone of a set of three-dimensional Helmholtz coils with its $\rm3\times3\, mm^{2}$ surfaces parallel to the $x$-coils, two pairs of $\rm3\times1.5\, mm^2$ surfaces parallel to the $y$-coils and $z$-coils, respectively, as shown in Fig.\,\ref{fig:setup}(f). The details of the setup are provided in the supplementary material. 
	
	According to the relative positions in our setup, a single $x$-axis magnetic field generated by the $x$-coils, which is aligned with $\left[001\right]$ or $\left[00\overline{1}\right]$ direction, is expected to equally project onto all four NV orientations. However, the practical result is apparently affected by the slight angle between the diamond and magnetic field. And this brings about a confounding combination of the four peaks without distinguishable hyperfine features in the ODMR spectrum. In addition, some uncertain factors in the practical setup can not be evaluated easily in advance, such as the unevenness of solidified optical adhesives and the angle error associated with mechanical structure. These make it hard to adjust the sample to the perfect aligned position manually. A better approach is to give a relatively strong main field $\vec B_{\rm main}$ ($x$-axis magnetic field here) with weak compensating field $\vec B_{\rm comp}$ (sum of $y$-axis and $z$-axis field here) to dynamically adjust the direction of the total bias field $\vec B_{\rm bias}=\vec B_{\rm main}+\vec B_{\rm comp}$. Though this approach is feasible, the dissimilar frequency shifts brought by the changes of magnetic field projections on four axes still make it time-consuming and non-optimum to adjust manually.  
	\begin{figure}
		\includegraphics{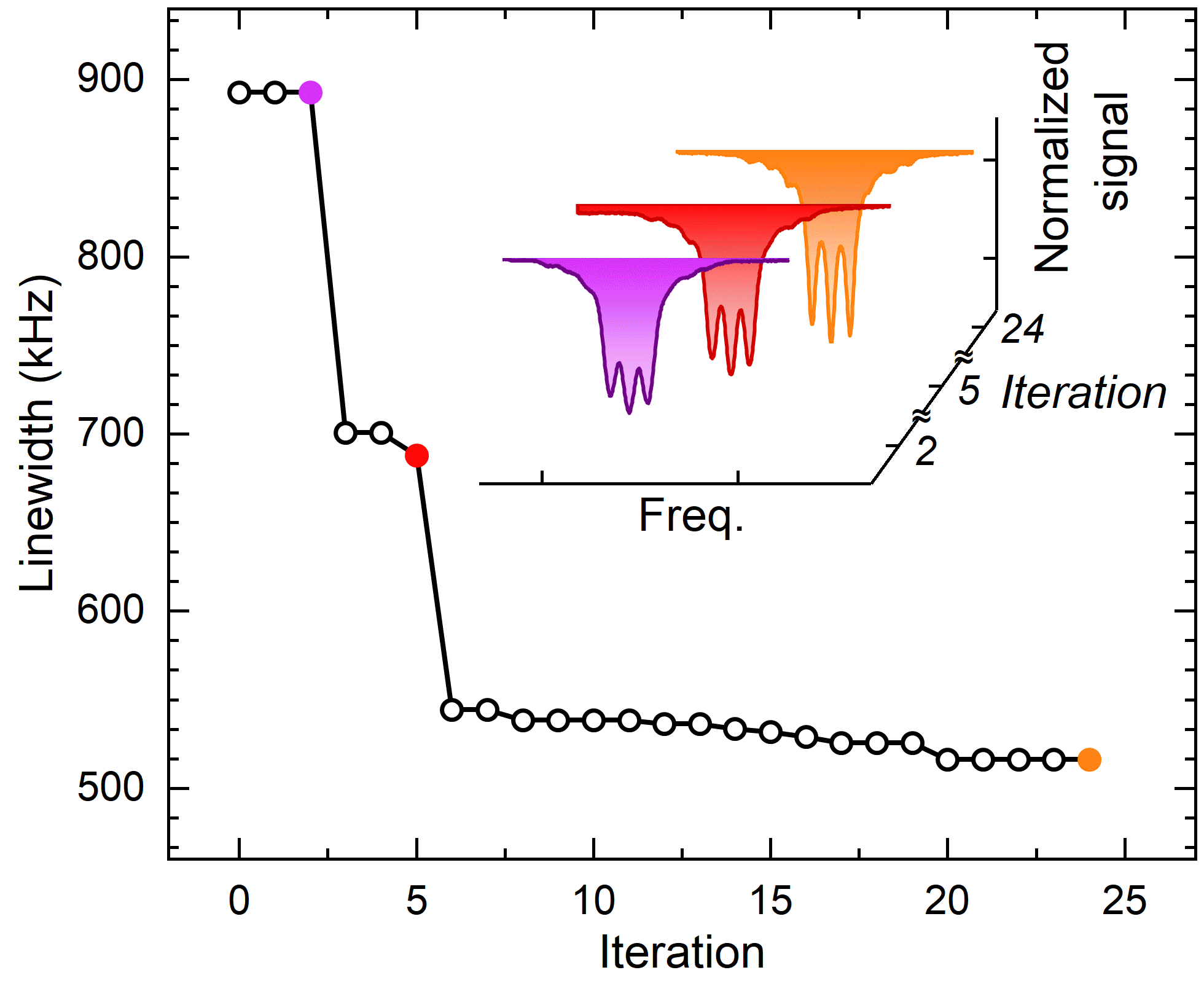}
		\caption{\label{fig:PSO}The results of reduced resonance linewidth with the number of algorithm iterations $\rm N_{iter}$ which is from 0 to 24. The inset shows the normalized cw-ODMR signals at the approximate microwave frequency of 2904 MHz varying as the convergence of the loss. The purple, red and orange points are located at the 2nd, 5th, and 24th iteration, respectively. Their linewidths and contrast ratios are $\rm 893\ kHz$, $1.55\ \%$, $\rm 688\ kHz$, $2.15\ \%$ and $\rm 516\ kHz$, $2.22\ \%$. To directly show the numerical changes, the variable corresponding to the vertical axis is set to the central linewidth instead of its reciprocal used as the loss function in the algorithm.}
	\end{figure} 
	
	\begin{figure*}
		\includegraphics{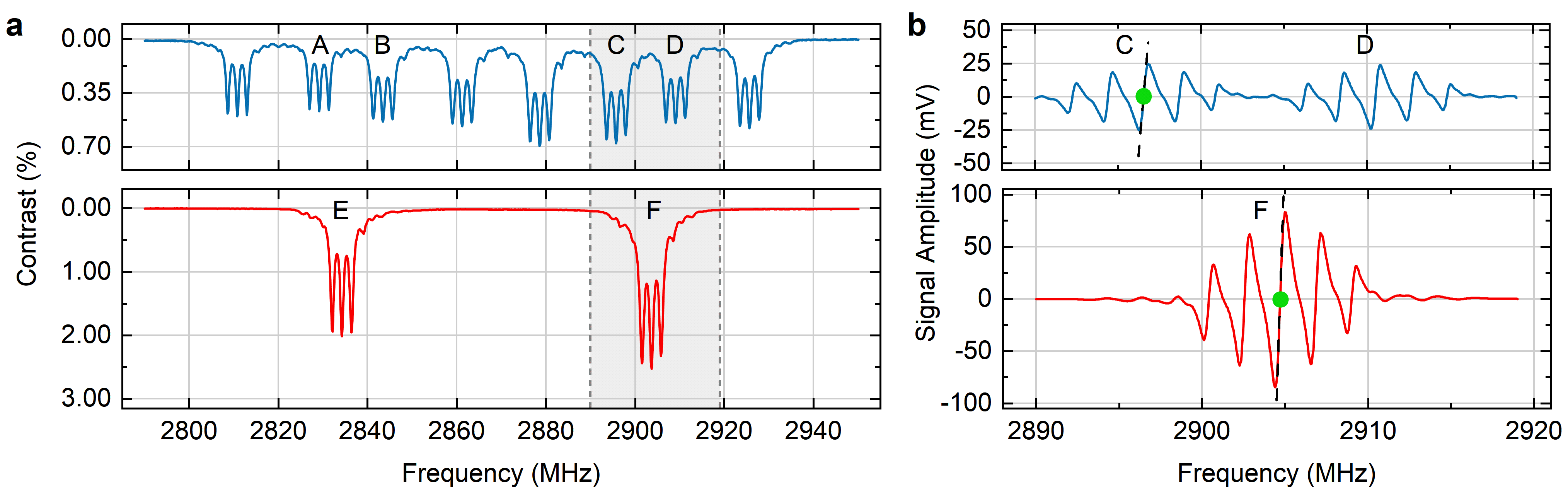}
		\caption{\label{fig:wide} (a) cw-ODMR spectra from an ensemble of NV centers in a bias magnetic field which causes complete separation of all the transitions (blue line) and another bias magnetic field (red line) which causes complete overlapping. (b) Frequency-modulated cw-ODMR spectra from $\rm2890\enspace MHz$ to $\rm2920\enspace MHz$ (marked in light gray in (a)). This frequency range includes group C, D and F. Each peak corresponds to five zero crossings due to the simultaneous resonance of all the three $^{14}$N hyperfine transitions. And the central zero crossing points utilized for magnetic sensing are marked in green for distinction.}
	\end{figure*}
	
	Here, we demonstrate the application of particle swarm optimization (PSO) algorithm into this processes. The PSO algorithm is a branch of evolutionary computation method and has been utilized in many fields including crystal structure prediction \cite{wang2010crystal} and quantum measurement \cite{hayes2014swarm}. PSO uses one set of ``particles", which contain two properties of position and velocity, to search the global optimal solution. Here, the optimization is in a 2-dimension space and  the particle position $\vec{x}$ is defined as
	\begin{equation}
		\label{eq:position define}
		\vec{x}=\left[B_{y},B_{z}\right]^T, 
	\end{equation}where $B_{y}$ and $B_{z}$ are numerical values of the y-axis and z-axis magnetic field, respectively. During the optimization process, paths of these particles will be affected by both local and global optimums. More generally, for the $i$th particle in an $N$-size swarm at the $k$th iteration, its local best location $\vec{l}_{i}^{k}$ can be determined by
	\begin{equation}
		\label{eq:l_best}
		f(\vec{l}_{i}^{k})=\mathop{max}\limits_{j\le k}(f(\vec{x}^{j}_i)),
	\end{equation}
	where $f(x)$ is the loss function to be maximized. And the global best location $\vec{g}^{k}$ can be determined by
	\begin{equation}
		\label{eq:g_best}
		f(\vec{g}^{k})=\mathop{max}\limits_{j\le k,0<i\le N}(f(\vec{x}^{j}_{i})).
	\end{equation}
	The update of this particle's velocity $\vec{v}^{k+1}_{i}$ and position $\vec{x}^{k+1}_{i}$ are carried out as
	\begin{equation}
		\label{eq:velocity update}
		\vec{v}^{k+1}_{i} = w\vec{v}^{k}_{i}+c_{l}\bm{r_{l}}\left(\vec{l}^{k}_{i}-\vec{x}^k_{i}\right)+c_{g}\bm{r_{g}}\left(\vec{g}^{k}-\vec{x}^{k}_{i}\right),
	\end{equation}
	\begin{equation}
		\label{eq:position update}
		\vec{x}^{k+1}_{i}=\vec{x}^{k}_{i}+\vec{v}^{k+1}_{i},
	\end{equation}
	where $w$ is the inertia weight, $c_{l}$ and $c_{g}$ are swarm factor and self-confidence factor respectively, $\bm{r_{l}}$ and $\bm{r_{g}}$ are 2-dimension diagonal matrices with the 2 non-zero elements independently and randomly generated from the numerical interval $\left(0, 1\right)$ at each iteration. As a whole, the new velocity of a particle are determined by three parts: its inertia of keeping the last velocity, history information from itself and the overall information from the swarm. The parameters $c_{l}$ and $c_{g}$ has been demonstrated to be the constant 2 in Ref. \cite{eberhart2000comparing,wang2010crystal}, which will be good for overall performance. According to this and for simplification, $w=1.0$, $c_{l}=c_{g}=2.0$ and $N=10$ are utilized throughout the iterative process. In addition, position and velocity of the particles are initialized with random values within boundary conditions, which is $\vec{x}_{min}\le\vec{x}\le\vec{x}_{max}$ and $\vec{v}_{min}\le\vec{v}\le\vec{v}_{max}$. These conditions are empirically decided in experiment to keep particles from leaving the searching area and moving too fast or slow.
		
	To measure the degree of overlapping, multiple-Lorentzians fitting is utilized and the number of peaks is set to the constant $3$ with their intervals in the range of $\rm2.16\times(1\pm10\%)\enspace MHz$.  The loss function here is selected as $f(x)=\rm 1/\Delta \nu$ instead of $f(x)=\mathcal{C}/\Delta \nu$ according to Eq.\,\ref{eq:sens}, where $\Delta \nu$ and $\mathcal{C}$ only refer to the central resonance linewidth of the three hyperfine features. This is reasonable due to the fact that the increase of $\mathcal{C}$ is accompanied by the decrease of $\Delta \nu$ in the process of transitions overlapping. The number of total iteration $\rm N_{iter}$ is not preset because the optimization process is continuous until convergence. Here, the criterion of convergence is based on the particles' trends towards the global best position on the 2-D mesh constructed from $B_{y}$ and $B_{z}$ with fixed $B_{x}=2.12\enspace \rm mT$. This means that the optimization is considered to converge when the density of particles reach a preset threshold value (see more details in supplement). The typical convergence curve with 10 particles used during 25 iterations is shown in Fig.\,\ref{fig:PSO}, where 250 ODMR sweeps are conducted and 40 minutes are spent. Though the time spent here is still relatively long, the global best field can be exactly given by this strategy finally, which is $\vec{B}_{\rm bias}\rm=(2.12, -0.016, -0.070)\enspace mT$. And we believe that this cost can be reduced by further improving the algorithm.
	
\section{MAGNETIC FIELD SENSITIVITY}	
	For comparison purpose, frequency domain sweeps using non-modulated microwave and frequency-modulated microwave are conducted under two different bias magnetic fields respectively. The signal from Monitor+ output of PD is acquired firstly with a home-built data acquisition system \cite{zhou2021mixed}. The spectrum above in Fig.\,\ref{fig:wide}(a) shows a completely separated ODMR features from $\rm2790\ MHz$ to $\rm2950\ MHz$ in $\vec{B}_{\rm bias}=\rm(2.12, -0.586, 1.01)\ mT$ with a step resolution of $\rm120\enspace kHz$. Inhomogeneous distribution of contrast is related to the $S_{11}$ trace of the microwave resonator applied here, which means a power uniformity in the frequency window \cite{ahmadi2017pump}. The subfeature at around $\rm2897\enspace MHz$ (the second peak of group C in gray-marked district) possesses a typical resonance linewidth of $\rm574\enspace kHz$ and a PL contrast of $\rm0.51\%$. In contrast, the spectrum below displays a complete transitions overlapping OMDR features in $\vec{B}_{\rm bias}\rm=(2.12, -0.016, -0.070)\enspace mT$ acquired in the previous section. And the subfeature (group F) at around $\rm2904\enspace MHz$ possesses a linewidth of $\rm550\enspace kHz$ and a contrast of $\rm2.04\%$. These parameters are slightly different from those shown in Fig.\,\ref{fig:PSO}(a) because of high average and fitting errors. In addition, complete transitions overlapping will enhance NV-NV interaction which may cause increase of resonance linewidth to some extent. However, we believe this difference has no significant effects on magnetic field sensitivity here (see supplementary material).
	
	In the following process, the microwave is frequency modulated at $\rm28\enspace kHz$, which is also the demodulation frequency of lock-in amplifier (LIA). According to previous works \cite{barry2016optical,ahmadi2017pump,patel2020subnanotesla}, simultaneously exciting all three hyperfine features shown in Fig.\,\ref{fig:wide}(a) is able to bring a contrast improvement to the central feature and result in a sensitivity enhancement. So the generated microwave is further mixed with a $\rm2.16\enspace MHz$ sinusoidal oscillation. The optimized microwave power is around $\rm-31\enspace dBm$ of each driving frequency and amplified by $\rm46\enspace dB$. Fig.\,\ref{fig:wide}(b) shows the signals generated from output X of the LIA whose profile is proportional to the first derivative of the ODMR lineshape approximately. Through linear fitting at zero-crossing points marked with green in Fig.\,\ref{fig:wide}(b), we determine their slopes of $\rm0.148\enspace \upmu V/Hz$ and $\rm0.476\enspace \upmu V/Hz$, which can be transformed into $\rm4.14\enspace \upmu V/nT$ and $\rm13.3\enspace \upmu V/nT$ by multiplying the ratio of $\gamma_{e}\sim \rm28\enspace MHz/mT$, respectively. As mentioned earlier, actual magnetic field in sensitivity-enhanced mode will be reduced by the factor of $\rm cos(\theta_{tet}/2)$. Hence, the central slope of the overlapped resonance utilized for sensitivity calculation should also be multiplied by this factor, and the final result we obtain is $\rm7.68\enspace \upmu V/nT$.

	\begin{figure}
	\includegraphics{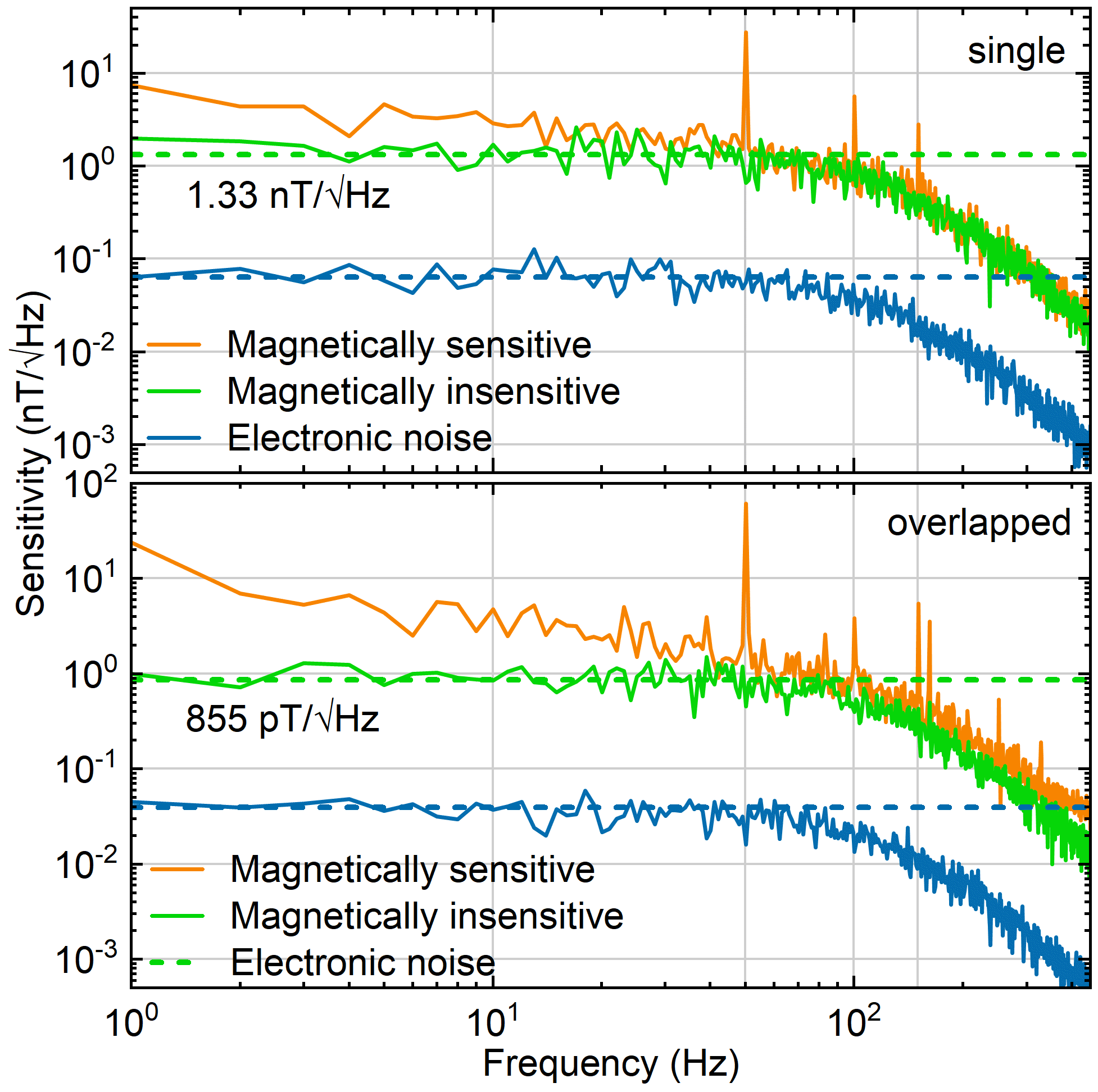}
	\caption{\label{fig:sensitivity}Magnetic field noise spectra utilizing the single resonance feature C (above) and the overlapped resonance feature F (below). The orange lines indicate the noise in magnetically sensitive configuration in our laboratory environment. The green lines indicate the noise in magnetically insensitive configuration. The blue lines indicate the electronic noise where the pump laser is turned off.}
	\end{figure}
	
	To give a direct demonstration of the enhancement, magnetic noise spectral density $\delta S(f)$ is measured under these two situations mentioned above and the sensitivities are calculated by $\nu_{B}(f)=\delta S(f)/\left|\partial S/\partial B\right|_{\rm max}$ \cite{ahmadi2017pump}, where $\left|\partial S/\partial B\right|_{\rm max}$ is the maximum slope of the lock-in amplified ODMR curve which has been obtained in previous section. The main results are summarized in Fig.\,\ref{fig:sensitivity}.
	
	As preparation for measurements, the time constant of LIA is set to $\rm1\ ms$ with a fourth-order low-pass filter, yielding a equivalent noise bandwidth of $\rm77.92\ Hz$. The microwave central frequency is set equal to the resonance frequency where the setup is magnetically sensitive. Then, the signal in 1 second is acquired at the sampling rate of $\rm53.57\ kSa/s$ and translated into magnetic field according to the slope calculated above. Fast-Fourier transform is utilized in data processing for noise analysis. Similar operations repeat when the microwave central frequency is set far away from the resonance point, where the setup is magnetically insensitive. The characteristic peaks at $\rm50\ Hz$, $\rm100\ Hz$ and $\rm150\ Hz$ are observed when the microwave is on resonance (orange lines in Fig.\,\ref{fig:sensitivity}) and disappear when the microwave is off resonance (green lines in Fig.\,\ref{fig:sensitivity}) in the spectra. These signals are attributed to ambient magnetic noise in the laboratory \cite{doi:10.1063/1.5095241,ahmadi2017pump}. 
	We directly average the magnetically insensitive signal from $\rm1$ to $\rm77.92\ Hz$. The voltage noise densities before and after overlapping within this bandwidth are $\rm5.50\ \upmu V/\sqrt{Hz}$ and $\rm 6.58\ \upmu V/\sqrt{Hz}$. Then, the resulting magnetic field sensitivities are further calculated as $\rm1.33\ nT/\sqrt{Hz}$ and $\rm855\ pT/\sqrt{Hz}$, which correspond to the separate and overlapping strategy, respectively. Thus, the reciprocal of the sensitivity is expanded by about $\rm1.5$ times. In addition, the original electronic noise floors acquired when the exciting laser is completely blocked are $\rm254\ nV/\sqrt{Hz}$ and $\rm265\ nV/\sqrt{Hz}$ respectively, which have no significant fluctuations. Therefore, we consider that the resulting electronic noise floors of $\rm61\ pT/\sqrt{Hz}$ and $\rm35\ pT/\sqrt{Hz}$ under two different bias fields are reasonable, but they are not commensurable with each other.

\section{CONCLUSION AND OUTLOOK}
	In this work, we demonstrate the utilization of four NV orientations for ensemble-NV magnetometry and the method of applying the optimal bias magnetic field. The complete overlapping of transitions corresponding to all NV orientations will bring a high improvement to the PL contrast and maximize the slope of central zero-crossing in lock-in amplified ODMR spectrum. Thus this results in an effective enhancement of magnetic field sensitivity. Moreover, we introduce PSO algorithm into the calibration procedure, which makes it automatic and relatively rapid to determine the optimal magnetic direction. With our setup, the sensitivity marked by the magnetic noise density is promoted from $\rm 1.33\enspace nT/\sqrt{Hz}$ to $\rm 855\enspace pT/\sqrt{Hz}$. 

	We believe that our approach, which is mainly based on dynamically changing the bias magnetic field, is easy to combine with other sensitivity enhancement methods. In the case of our setup, there is a lot of room for enhancement in collection efficiency by using a compound parabolic concentrator (CPC) \cite{wolf2015subpicotesla} or a total internal reflection (TIR) lens \cite{xu2019high}. Furthermore, we expect that more other variables, such as excitation power and microwave power, can also be introduced into the calibration process to reach the global best ratio of the resonance linewidth to PL contrast. And this will be greatly helpful in exploring the potential optimal sensitivity of an ensemble-NV magnetometer.          

	\section*{SUPPLEMENTARY MATERIAL}
	See the supplementary material for more details of the setup, the algorithm and the data analysis.
	~\\
	
	\section*{Acknowledgments}
	This work is supported by the National Key R\&D Program of China (Grant Nos. 2020YFA0309400 and 2018YFA0306600), the National Natural Science Foundation of China (Grant Nos. 12174081, 11904070), the Fundamental Research Funds for the Central Universities (Grant Nos. JZ2021HGTB0126 and PA2021KCPY0052), and Special project A of scientific research and innovation for young teachers of Hefei University of Technology (Grant No. JZ2020HGQA0165).
	
	\section*{AUTHOR DECLARATIONS}
	\subsection*{Conflict of interest}
	The authors have no conflicts to disclose.
	
	\section*{Data Availability Statement}
	The data that supports the findings of this study are available within the article and its supplementary material upon reasonable request.

\bibliography{main}
	
\end{document}